\documentclass[12pt]{article}
\usepackage{amssymb}
\date{ }
\title{ Quantum-like approach to financial risk: quantum anthropic
principle
} {\Huge \bf \author{E. W. Piotrowski\\ Institute of Theoretical
Physics, University of Bia\l ystok,\\ Lipowa 41, Pl 15424 Bia\l
ystok, Poland\\ e-mail: ep@alpha.uwb.edu.pl\\ and J. S\l adkowski
\\ Institute of Physics, University of Silesia, \\ Uniwersytecka
4, Pl 40007 Katowice, Poland \\ e-mail: sladk@us.edu.pl}}
\begin{document}
\maketitle
\def\Z{{\bf Z\!\!Z}}
\def\R{{\bf I\!R}}
\def\N{{\bf I\!N}}
\def\C{{\mathbb C}}
\begin{abstract}
We continue the analysis of quantum-like description of market
phenomena and economics. We show that it is possible to define a
risk inclination operator acting in some Hilbert space that has a
lot of common with quantum description of the harmonic oscillator.
The approach has roots in the recently developed quantum game
theory and quantum computing. A quantum anthropic principle is
formulated

\end{abstract}
\section{Noncommutative quantum mechanics}
The term {\it noncommutative quantum mechanics} is misleading as
it suggest there might exists something like commutative quantum
mechanics. In fact quantum theory has built in various types of
noncommutativity since its infancy. Nowadays the adjective
noncommutative reflects the possibility that the space variables
$\hat{x}^i$ might not commute. This sort of generalizations has it
roots in M-theory (branes), deformation quantization, Weyl
quantization and  curiosity \cite{1}. This is of course connected
with the A. Connes program \cite{2,3,4}. Among various possible
generalizations the most popular one are
$$ [\hat{x}^i,\hat{x}^j] = i \theta^{ij} ,  \;\theta^{ij} \in \C,
\eqno(1a)$$
$$[\hat{x}^i,\hat{x}^j] = i C^{ij}{}_k \hat{x}^k ,\;  C^{ij}{}_k \in \C,
\eqno(1b)$$
$$ \hat{x}^i \hat{x}^j = {\mathrm q}^{-1}\hat R^{ij}{}_{kl}
\hat{x}^k \hat{x}^l , \;\hat{R}^{ij}{}_{kl} \in \C . \eqno(1c)$$
\noindent In all these cases the index $i$ takes  values from $1$
to $N$. We shall suppose that the appropriate algebra $A_x$ has a
unit element \cite{2,3}. Eqs.~(1b) and (1c) describe the so called
Lie algebraic and quantum group generalizations, respectively.
Here we would be interested in a generalization of the form given
by Eq.~(1a) which may traced back to works of H. Weyl \cite{5} and
Moyal \cite{16}.
\section{Quantum market games}
It was a complete surprise to us to find the University of Bologna
preprint entitled {\it Quantum Mechanics and Mathematical
Economics are Isomorphic} and written by Lambertini \cite{6}. We
were actually working on the direct application of recently
developed quantum game theory \cite{7,8,9} in economics \cite{10}
- \cite{15}. The Lambertini's paper  was a further incentive to
work harder. In the "standard" quantum game theory one tries in
some sense to quantize an operational description of "classical"
versions of the game being analyzed. It usually enlarges  the set
admissible strategies in a nontrivial way. We follow a different
rout. The market players strategies are described in terms state
vectors $|\psi\rangle$ belonging to some Hilbert space ${\cal H}$
\cite{11,12,14}. The probability densities of revealing the
players, say Alice and Bob, intentions are described in terms of
random variables $p$ and $q$:
\begin{equation}
\label{eigenstosc} \frac{|\langle q|\psi\rangle_A|^2}{\phantom{}_A
\langle\psi|\psi\rangle_A}\, \frac{|\langle
p|\psi\rangle_B|^2}{\phantom{}_B \langle\psi|\psi\rangle_B}\;d q d
p\, ,
\end{equation}
where $\langle q|\psi\rangle_A$\/ is the probability amplitude of
offering the price $q$ by Alice who wants to buy and the demand
component of her state is given by
$|\psi\rangle_A\in\mathcal{H}_{A}$\/. Bob's amplitude $\langle
p|\psi\rangle_B$\/ is interpreted in an analogous way (opposite
position). Of course, the "intentions" $q$ and $p$ not always
result in the accomplishment of the transaction \cite{12}. If one
considers the following facts\cite{10,  17,18}:
\begin{itemize}
\item error theory: second moments of a random
variable describe errors
\item M. Markowitz's portfolio theory (Nobel Prize 1990)
\item L. Bachelier's theory of options:  the random variable $q^{2} + p^{2}$ measures joint risk
for a stock buying-selling transaction ( Merton \& Scholes won
Nobel Prize in 1997)
\end{itemize}
then it seems reasonable to define the observable {\it the risk
inclination operator}:
$$
H(\mathcal{P}_k,\mathcal{Q}_k):=\frac{(\mathcal{P}_k-p_{k0})^2}{2\,m}+
                     \frac{m\,\omega^2(\mathcal{Q}_k-q_{k0})^2}{2}\,,
\eqno(2) $$ \noindent where $p_{k0}:=\frac{
\phantom{}_k\negthinspace\langle\psi|\mathcal{P}_k|\psi\rangle_k }
{\phantom{}_k\negthinspace\langle\psi|\psi\rangle_k}\,$,
$q_{k0}:=\frac{
\phantom{}_k\negthinspace\langle\psi|\mathcal{Q}_k|\psi\rangle_k }
{\phantom{}_k\negthinspace\langle\psi|\psi\rangle_k}\,$,
$\omega:=\frac{2\pi}{\theta}\,$.  $ \theta$ denotes the
characteristic time of transaction \cite{10} which is, roughly
speaking, an average time spread between two opposite moves of a
player (e.~g.~buying and selling the same asset). The parameter
$m>0$ measures the risk asymmetry between buying and selling
positions. Analogies with quantum harmonic oscillator allow for
the following characterization of quantum market games. The
constant $h_E$ describes the minimal inclination of the player to
risk. It is equal to the product of the lowest eigenvalue of
$H(\mathcal{P}_k,\mathcal{Q}_k) $ and $2\theta$. $2\theta $ is in
fact the minimal interval during which it makes sense to measure
the profit.  Except the ground state all the adiabatic strategies
$H(\mathcal{P}_k,\mathcal{Q}_k)|\psi\rangle={const}|\psi\rangle$
are giffens \cite{10, 12} that is goods that do not obey the law
of demand and supply. It should be noted here that in a general
case the operators $\mathcal{Q}_k $ do not commute because traders
observe moves of other players and often act accordingly. One big
bid can influence the market at least in a limited time spread.
Therefore it is natural to apply the formalism of noncommutative
quantum mechanics where one considers
$$ [ x^{i},x^{k}] = i \Theta ^{ik}:=i\Theta \,\epsilon ^{ik}.\eqno(3) $$
The analysis of harmonic oscillator in more then one dimensions
\cite{18} imply that the parameter $\Theta $ modifies the constant
$\hslash_E$ $\rightarrow \sqrt{\hslash_E^{2} + \Theta ^{2}} $ and,
accordingly, the eigenvalues of $H(\mathcal{P}_k,\mathcal{Q}_k)$.
This has the natural interpretation that moves performed by other
players can diminish or increase one's inclination to taking risk.
\section{ Market as a measuring apparatus} When a game allows a
great number of players in it is useful to consider it as a
two-players game: the trader $|\psi\rangle_{k}$ against the Rest
of the World (RW). The concrete algorithm $\mathcal{A}$ may allow
for an effective  strategy of RW (for a sufficiently large number
of players the single player strategy should not influence on the
form of the RW strategy). If one considers the RW strategy it make
sense to declare its simultaneous demand and supply states because
for one player RW is a buyer and for another it is a seller. To
describe such situation it is convenient to use the Wigner
formalism. The following subsection  describe shortly various
aspects of quantum markets.
\subsection{ Quantum Zeno effect} If the market  continuously
measures the same strategy of the player, say the demand $\langle
q|\psi\rangle $, and the process is repeated sufficiently often
for the whole market, then the prices given by some algorithm  do
not result from the supplying strategy $\langle p|\psi\rangle $ of
the player. The necessary condition for determining the profit of
the game is the transition of the player to the state $\langle
p|\psi\rangle $. If, simultaneously, many of the players changes
their strategies then the quotation process may collapse due to
the lack of opposite moves. In this way the quantum Zeno \cite{20}
effects explains stock exchange crashes.  Another example of the
quantum market Zeno effect is the stabilization of prices of an
asset provided by a monopolist.
\subsection{ Eigenstates of
$\mathcal{Q}$ and $\mathcal{P}$} Let us suppose that the
amplitudes for the strategies $\langle q|\psi\rangle_{k}$ or
$\langle p|\psi\rangle _{k}$ that have infinite integrals of
squares of their modules, ($\langle q|\psi\rangle_k\not\in L^2$)
have the natural interpretation as the will of the $k$-th player
of buying (selling) of the amount $ d_k$ ($s_k$) of the asset
$\mathfrak{G}$. So the strategy $\langle q|\psi\rangle_k= \langle
q|a\rangle =\delta(q,a)$ means that in the case of classifying the
player to the set $\{k_d\}$, refusal of  buying cheaper than at
$c=e^{a}$ and the will of buying at any  price equal or above
$e^{a}$. In the case of "measurement" in the set $\{k_d\}$ the
player declares the will of selling at any price. The above
interpretation is consistent with the Heisenberg uncertainty
relation. The strategies $\langle q|\psi\rangle_2=\langle
q|a\rangle$ (or $\langle p|\psi\rangle_2=\langle p|a\rangle$) do
not correspond to the RW behaviour because the conditions
$d_2,s_2>0$, if always fulfilled, allow for unlimited profits (the
readiness to buy or sell $\mathfrak{G}$ at any price). The demand
and supply functions give probabilities of coming off transactions
in the game when the player use the strategy $\langle
p|{const}\rangle$ or $\langle q|{const}\rangle$ and RW, proposing
the price, use the strategy $\rho$.
\subsection{ Correlated coherent strategies} We will define
correlated coherent strategies as the eigenvectors of the
annihilation operator $\mathcal{C}_k$
$$
\mathcal{C}_k(r,\eta):=\frac{1}{2\eta}\Bigl(1+\frac{ir}{\sqrt{1-r^2}}
\Bigr)\mathcal{Q}_k + i\eta\mathcal{P}_k , \eqno(4)
$$
where $r$ is the correlation coefficient  $r\in[-1,1]$, $\eta>0$.
In these strategies buying and selling transactions are correlated
and the product of dispersions fulfills  the Heisenberg-like
uncertainty relation
$\Delta_p\Delta_q\sqrt{1-r^2}\geq\frac{\hslash_E}{2}$ and is
minimal. The annihilation operators $\mathcal{C}_k$ and their
eigenvectors may be parameterized by
$\Delta_p=\frac{\hslash_E}{2\eta}\,$,
$\Delta_q=\frac{\eta}{\sqrt{1-r^2}}$ and $r$.
\subsection{ Mixed
states and thermal strategies} According to classics of game
theory  the biggest choice  of strategies is provided by the mixed
states $\rho(p,q)$. Among them the most interesting are the
thermal ones. They are characterized by constant inclination to
risk, $E(H(\mathcal{P},\mathcal{Q}))={const}$ and maximal entropy.
The Wigner measure for the $n$-th exited state of harmonic
oscillator have the form
$$
W_n(p,q)dpdq=\frac{(-1)^n}{\pi\hslash_E}\,\thinspace
e^{-\frac{2H(p,q)}{\hslash_E\omega}}
\,L_n\bigl(\frac{4H(p,q)}{\hslash_E\omega}\bigr)\,dpdq, \eqno(5)
$$ where $L_{n}$ is the $n$-th Laguerre polynomial. The mixed state  $\rho_\beta$
determined by the Wigner measures $W_ndpdq$ weighted by the Gibbs
distribution $w_n(\beta):=\frac{e^{-\beta
n\hslash_E\omega}}{\sum_{k=0}^\infty e^{-\beta k\hslash_E\omega}}$
have the form
\begin{eqnarray*}
\rho_\beta (p,q)dpdq:&=&\sum_{n=0}^\infty w_n(\beta) W_n(p,q) dpdq\\
&=&\frac{\omega}{2\pi}\;x\; e^{-xH(p,q)} \Bigr|_{x=
\frac{2}{\hslash_E\omega}\tanh(\beta\frac{\hslash_E\omega}{2})}
dpdq .\end{eqnarray*}
\subsection{ Market cleared by quantum
computer} When the algorithm $\mathcal{A}$ calculating in a
separable Hilbert space $H_k$ does not know the players strategies
it must choose the basis in an arbitrary way. This may result in
arbitrary long representations of the amplitudes of strategies.
Therefore the algorithm $\mathcal{A}$ should be looked for in the
NP (non-polynomial) class and quantum markets may be formed
provided the quatum computation technology is possible. Is the
quantum arbitrage possible only if there is a unique
correspondence between $\hslash$ and $h_E$?  Was the hypothetical
evidence given by Robert Giffen in the British Parliament the
first ever description of quantum phenomenon?  The commonly
accepted universality of quantum theory should encourage physicist
in looking for traces quantum world in social phenomena.
\section{Conclusions}
We think that the formalism of quantum theory may provide us with
tools of unexpected power that combined with methods of game
theory may allow for much deeper understanding of financial
phenomena than it is usually expected. Let us quote the
 Editor's Note to Comlexity Digest 2001.27(4)
(http://www.comdig.org) "It might be that while observing the due
ceremonial of everyday market transaction we are in fact observing
capital flows resulting from quantum games eluding classical
description. If human decisions can be traced to microscopic
quantum events one would expect that nature would have taken
advantage of quantum computation in evolving complex brains. In
that sense one could indeed say that quantum computers are playing
their market games according to quantum rules". During the past
decade options gained a significant position in capital turnover
all over the world. It may mean that simple methods of
maximization of profit became unsatisfactory. At present
minimization of financial risk is playing the key role. Lower risk
means better prognosises for industrial and market development and
this results in higher profits. Methods minimazing the financial
risk will become a focus of investors attention.  All this tempt
us into formulating the {\em quantum anthropic principle} of the
following form. {\em At earlier civilization stages markets are
governed by classical laws (as classical logic prevailed in
reasoning) but the incomparable efficacy of quantum algorithms in
multiplying profits \footnote{Note the significance of quantum
phenomena in modern technologies and their influence on economics}
will result in quantum behaviour prevailing over the classical
one}. We envisage markets cleared by quantum algorithms
(computers) \cite{10, 21}.

\end{document}